# Electrophysiological Investigation of Insect Pain Threshold

*[1]Marc Josep Montagut Marques, *[2]Pan Minghao, [3]Ryuichi Okada, [3]Midori Sakura, **[4]Kayo Hirose, **[1,2]Shinjiro Umezu

[1]Department of Integrative Bioscience and Biomedical Engineering, Waseda University, Tokyo, Japan

[2]Department of Modern Mechanical Engineering, Waseda University, Tokyo, Japan

[3]Department of Biology, Graduate School of Science, Kobe University, Kobe, Japan

[4]Department of Anesthesiology and Pain Relief Center, The University of Tokyo Hospital, Tokyo, Japan

*First author

**Corresponding authors

## 0. Abstract

The question of whether insects experience pain has long been debated in neuroscience and animal behavior research. Increasing evidence suggests that insects possess the ability to detect and respond to noxious stimuli, exhibiting behaviors indicative of pain perception. This study investigates the relationship between pain stimuli and physiological responses in crickets (Gryllidae), focusing on heart rate (ECG) and brain wave (EEG) patterns. We applied a range of mechanical, chemical, thermal, and electrical stimuli to crickets, recording ECG and EEG data while employing a deep learning-based model to classify pain levels. Our findings revealed significant heart rate changes and EEG fluctuations in response to various stimuli, with the highest intensity stimuli inducing marked physiological stress. The AI-based analysis, utilizing AlexNet for EEG signal classification, achieved 90% accuracy in distinguishing between resting, low-pain, and high-pain states. While no social sharing of pain was observed through ECG measurements, these results contribute to the growing body of evidence supporting insect nociception and offer new insights into their physiological responses to external stressors. This research advances the understanding of insect pain mechanisms and demonstrates the potential for AI-driven analysis in entomological studies.

## 1. Introduction

The question of whether insects experience pain has long been a subject of debate in neuroscience and animal behavior research. Increasing evidence suggests that insects possess the ability to detect and respond to noxious stimuli, exhibiting behaviors indicative of pain perception [1]. Many species demonstrate rapid escape responses when exposed to harmful mechanical, thermal, or chemical stimuli. For example, crickets (*Gryllidae*) exhibit swift avoidance behaviors when subjected to heat or mechanical touch, suggesting an intrinsic ability to detect and respond to potentially harmful stimuli [2]. Furthermore, fruit flies (*Drosophila melanogaster*) have been observed to develop heightened sensitivity to noxious stimuli following injury, a phenomenon analogous to chronic pain states in vertebrates [3].

A comprehensive meta-analysis of over 300 studies by Queen Mary University of London further supports the argument that insects exhibit behavioral responses consistent with pain perception [4]. These behaviors include self-protective reactions, learned avoidance, and prolonged hypersensitivity following tissue damage, all of which mirror pain-associated responses in vertebrates. Additionally, research has highlighted the presence of specialized nociceptive sensors in insects and other invertebrates, which detect tissue damage and other harmful factors that trigger avoidance behaviors, further reinforcing the possibility of pain perception in these organisms [5, 6].

From a neurophysiological perspective, studies have revealed that insects generate distinct neural signals in response to noxious stimuli. Electrophysiological investigations have identified neurons extending from the brain to the ventral nerve cord that are functionally analogous to vertebrate nociceptive pathways, playing a key role in insect pain-like responses [7, 8]. Moreover, molecular studies indicate that insects' express neurotransmitters and proteins involved in pain processing, such as transient receptor potential (TRP) channels, which are conserved across multiple taxa and are known mediators of nociceptive signaling in vertebrates [9-11].

Despite accumulating evidence on behavioral and neurophysiological indicators of pain in insects, the relationship between pain perception and autonomic physiological responses remains poorly understood. In mammals, exposure to painful stimuli leads to significant changes in heart rate, mediated by the autonomic nervous system via sympathetic and parasympathetic pathways [12]. However, insects lack a conventional autonomic nervous system, relying instead on neurohormonal control mechanisms for cardiac regulation. Environmental factors such as temperature and humidity, as well as mechanical and chemical stimuli, are known to influence insect heart rate, yet it remains unclear whether pain-like experiences in insects elicit similar physiological responses as in mammals [13]. AI has demonstrated the feasibility of pain simulations to condition the human brain [14]. However, further data is needed, and test subjects remain limited. Insects offer a vast dataset for exploring the complexities of brain communication signals [15].

This study aims to address this gap by investigating the correlation between pain stimuli and cricket physiological response. By applying noxious stimuli of varying intensities while simultaneously recording electrocardiograms (ECGs) and electroencephalograms (EEGs), we seek to determine whether insects exhibit pain-related cardiac responses through the use of machine learning by developing an AI curated on the observation of gradual pain levels. Our findings will contribute to the ongoing debate on insect nociception, providing crucial physiological insights that may have implications for both fundamental neuroscience and applied entomology, including pest management and insect welfare considerations.

## 2. Methods

### 2.1 Pain application

A range of stimuli was applied to assess the physiological responses of crickets under controlled conditions. Light stimulation was delivered using an adjustable light source, allowing for the simulation of weak ambient light fluctuations. Mechanical pressure was administered at three intensity levels—low, moderate, and high—by varying both the contact area and the applied force. Wind stimulation was introduced through a controlled air blower, directing low, moderate, and high wind intensities toward the cricket's face. Similar wind-based stimuli have been used in studies of cricket cercal responses to air currents, where neuronal activity was recorded to assess sensitivity to different stimulus intensities [16]. Thermal stimuli were applied using an adjustable heat source and a high-temperature lighter, carefully regulating the temperature at the stimulation site to achieve low, moderate, and high levels of heat exposure. Research on *Drosophila melanogaster* has utilized similar thermal nociception assays to evaluate pain-like responses in fruit fly larvae [17]. Chemical stimulation was conducted using both volatile and contact-based methods. Weak-smelling volatile chemicals and mild coating agents were used to simulate low-intensity environmental exposure, while high-intensity stimulation involved strong volatile substances and direct application of concentrated reagents. All chemical dosages were strictly controlled to be non-lethal while eliciting a clear stress response. A comparable approach was employed in bumblebee studies to investigate motivational trade-offs in nociception, where bees were exposed to noxious chemical stimuli while making foraging decisions [18].

Electrical stimulation was applied using a precision electrical stimulator capable of delivering stable low- and high-intensity currents, ensuring safe and consistent electrode contact with specific regions of the cricket's body. Crickets were placed in a custom-designed observation chamber, where environmental conditions were maintained to minimize external interference. Stimuli were applied sequentially according to a predetermined order of intensity levels. Each stimulus, except for chemical stimulation, was delivered for one second, with a five-minute interval between applications to allow the crickets to recover to baseline conditions and prevent cumulative effects. Chemical stimuli were administered over a one-hour period to assess the effects of prolonged exposure. Each stimulation intensity was repeated three times per individual, and multiple crickets were tested to ensure a sufficient sample size and reduce variability due to individual differences. This experimental protocol was designed to ensure precise stimulus delivery while minimizing confounding variables that could affect physiological responses.

## 2.2 ECG sensor detection for multiple crickets

The ECG device was constructed using an AD8232 single-lead heart rate monitor. To accommodate the small size of the crickets, the standard electrodes were replaced with 200-micron acupuncture needles. The front-end chip configuration remained in its default setting to ensure signal stability. Data acquisition was performed by logging the recorded signals into a computer for subsequent analysis.

The crickets that have adapted to the environment are randomly divided into multiple groups, with four crickets in each group. During the experiment, the four crickets are first fixed in a 15 cm square shape with each insect staying in one of each corner, and then the ECG sensor is installed. At the same time, the ECG detection device is turned on, and the behavioral characteristics and ECG signals of the crickets are observed synchronously. After the cricket's physical signs are stable and the ECG signal is stable, the ECG signal of the cricket in the resting state is recorded in real time. After that pain is inflicted to one of the insects the reaction is observed on the others to understand whether there are social implications in the feeling of pain.

2.3 EEG experiment and AI analysis

Crickets were randomly assigned to either an experimental group or a control group, ensuring that each group contained multiple individuals to enhance statistical robustness. The experimental group was subjected to external stimulation treatments, whereas the control group remained in a resting state to facilitate comparative analysis.

Two types of stimuli were applied to the crickets in the experimental group. First, Chemical Coating Stimulation was administered by applying a predetermined concentration and dose of a chemical stimulus using a controlled dispensing device. EEG fluctuations were continuously monitored throughout the stimulation period to assess physiological responses. Second, High-Intensity Temperature Stimulation involved administering a brief high-temperature stimulus to evaluate acute thermal stress responses. EEG signals were recorded before and after stimulation to capture dynamic physiological changes. To prevent cumulative effects and ensure recovery to baseline, a minimum interval of 10 minutes was maintained between successive stimuli. Multiple crickets were included in each trial to enhance sample size reliability and mitigate the influence of individual variability.

EEG signals were recorded under both resting and stimulated conditions using a modified HARU-2 biosensor. Each data collection session was limited to five minutes to ensure sufficient data acquisition without imposing excessive stress on the specimens. The analog signals were converted into digital data via an acquisition card and subsequently stored for further analysis. A Fast Fourier Transform (FFT) algorithm was implemented to perform preliminary signal analysis, evaluating data quality and eliminating artifacts such as noise interference and anomalous values. Several hundred high-quality data segments were selected as the basis for subsequent machine learning analysis.

A deep learning-based neural network model was developed to analyze the processed EEG data. The model was trained using labeled datasets corresponding to both resting and stimulated conditions, with the objective of distinguishing between high-pain, low-pain and non-pain states in crickets. Multiple rounds of training and validation were conducted to refine the model's accuracy and improve its ability to detect subtle variations in EEG associated with nociceptive responses.

## 3 Results and discussion

### 3.1 Classification of Pain Levels

The results demonstrated significant differences in heart rate responses and behavioral reactions between the control and experimental groups. Crickets subjected to chemical coating stimulation exhibited immediate increases in heart rate, indicating heightened physiological activity. In contrast, those exposed to high-intensity temperature stimulation displayed abrupt heart rate spikes followed by gradual declines, suggesting acute stress responses. The intervals between stimuli ensured the recovery of baseline heart rate levels, reducing the potential for cumulative stress effects and allowing for accurate comparative assessments.

Behavioral responses were categorized into five levels based on stimulus intensity. **Level 1** stimuli, such as low light or exposure to chemical odors, resulted in slight reactions, including minor directional changes or movement. **Level 2** stimuli, including low oppressive forces, mild wind, and flashing light, triggered brief escape responses that ceased quickly. **Level 3** involved moderate thermal exposure, mid-level wind and pressure, and chemical coating, leading to more pronounced escape behaviors that persisted for a short duration before stopping. **Level 4** included middle-intensity thermal and electrical stimuli, combined with strong wind and pressure, eliciting prolonged and vigorous escape reactions. **Level 5**, characterized by high-intensity electrical and thermal stimulation along with exposure to strong chemical odors and coatings, induced behaviors resembling sustained struggle, with continued movement over an extended period. The response level and the overview of the reaction is summarized in Fig. 1.

Signal processing and machine learning analysis further corroborated these findings. The FFT algorithm effectively filtered noise and isolated key signal components, providing a robust dataset for neural network training. The deep learning model successfully classified pain and non-pain states with high accuracy, revealing distinct patterns in heart rate variability under different stimulation conditions. These findings highlight the potential of AI-driven analysis in entomological research, offering new insights into how insects physiologically respond to external stressors.

### 3.2 Social pain transfer

Comparative experiments demonstrated that the improved ECG sensor designed for crickets significantly enhanced signal collection quality. Under resting conditions, the sensor successfully captured a clear and typical ECG waveform of crickets, validating the rationality and effectiveness of its design. These findings confirm the feasibility of utilizing the sensor for cricket ECG detection, ensuring its reliability for further physiological studies.

Through optimization of both the experimental device and data acquisition program, minimal-invasive ECG detection of four crickets simultaneously was achieved by using fine acupuncture needles as seen in Fig 2a. Electrocardiogram data were synchronously collected and recorded on the same time axis, allowing for parallel monitoring of heart rate variations among individuals within a group. This advancement not only improves experimental efficiency but also establishes a technical platform for studying inter-cricket interactions. For instance, it enables the observation of whether crickets exhibit stress responses when one individual is subjected to pain stimulation, providing valuable insights into their physiological and behavioral dynamics.

The synchronous ECG recordings did not reveal any clear interactions between crickets as seen in Fig. 2b and c in response to pain stimuli. There were no observable changes in the heart rate signals of unaffected crickets when one individual experienced a painful stimulus. While this suggests a lack of physiological response to social distress, the results do not definitively confirm whether crickets perceive or acknowledge social harm. Since ECG measurements primarily serve as indicators of physiological stress and direct physical pain, they may not fully capture potential behavioral or neural responses to social stimuli. Additionally, no physical reactions were recorded in the unaffected crickets, further indicating an absence of observable social distress responses in this experimental setup.

Despite these advancements, the current technology still exhibits notable limitations. During long-term monitoring, data accuracy decreases as experiment time increases, leading to waveform distortion and potential errors in ECG readings. Compared to high-precision human ECG detection standards, the resolution and accuracy of the cricket ECG detection system require further improvement. In particular, the system struggles to capture minor fluctuations in heart rate, making it difficult to accurately quantify physiological responses to weak stimuli. These limitations hinder the in-depth exploration of fine-scale physiological changes in crickets and highlight the need for further technological refinement.

### 3.3 AI assisted pain categorization based on EEG signal

Fine electrodes were installed in the crickets' head area in order to reach brain wave signals as seen in Fig 3a and b. The insect was then subjected to pain conditions based on our level classification. Three classifications were selected in order to train the AI model, relaxed, low-damage and high damage based on level. For the EEG tests chemical pain was chosen as it does not affect the insect physically avoiding ambient artifacts in the signal. Significant fluctuations in EEG frequency are observed multiple times during the stimulation period, with abnormal cycles closely linked to the action mechanism of the chemical substance and the cricket's physiological stress response. These findings suggest that chemical stimulation induces prolonged physiological stress, leading to sustained alterations in the EEG signal.

For comparison the insect was subjected to temperature pain, a brief (1s) high-intensity temperature stimulus causes a sharp but short-term change in the cricket's EEG rate signal. The signal rate frequency increases instantaneously, with peak values reaching several times that of the resting state before gradually returning to near-baseline levels. This response indicates that crickets exhibit a strong immediate physiological reaction to sudden high-temperature exposure. However, the EEG system demonstrates a degree of self-regulation, allowing for a relatively rapid recovery to stability following the stimulus.

Following extensive data training and optimization, the AI model demonstrated high accuracy in determining whether a cricket was experiencing a pain state. The model was generated with AlexNet [19] and achieved an accuracy rate of 90% on the test set (specific accuracy may vary based on actual verification results), effectively distinguishing heart rate characteristics between resting and two levels of stimuli. The data provided to the model was the FFT of the three sensing channels of the portable EEG device. Fig 3c shows an example of resting data used for the training of the model. Fig 3d are use cases of the model to classify the different states of the cricket based on EEG data. These results highlight the potential of AI-driven analysis as a powerful tool for further investigations into cricket pain perception and physiological responses to external stimuli.

## 4 Conclusions

This study has yielded significant findings in the investigation of insect pain perception and physiological mechanisms. The pain level classification system, developed based on cricket behavioral responses, effectively establishes the relationship between various stimulation methods, intensities, and pain perception. This framework provides a foundation for further exploration into the physiological mechanisms underlying insect pain. From a technological perspective, advancements in sensor development, program implementation, and experimental device construction have led to breakthroughs in cricket ECG detection. The successful validation of a self-developed sensor and multi-cricket synchronous detection technology serves as a robust platform for future research. Additionally, the integration of a portable EEG instrument with AI-based analytical methods has facilitated the identification of cricket EEG patterns and enabled the establishment of a pain assessment model.

Despite these advancements, certain limitations remain. In ECG detection, challenges persist in achieving long-term and high-precision measurements. Additionally, in studies combining the portable EEG instrument with AI, the invasive nature of the detection method may affect data accuracy. Furthermore, the AI model still requires substantial improvements to enhance its precision in assessing pain levels.

Overall, this study provides new insights and directions for research on insect pain perception. While certain technical and methodological limitations exist, the findings establish a solid foundation for future investigations. Moving forward, further refinement of the pain classification system, advancements in detection technology, and optimization of AI models will be essential to advancing insect physiology research. These efforts may also contribute valuable insights into the broader understanding of pain mechanisms, including potential implications for human pain studies.

## 7. Figure legends

Figure 1

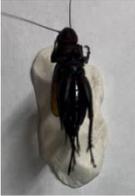

**Pain Level Classification in Crickets.** Five pain levels were established based on the crickets' reactions to various stimuli, each targeting different areas of the nervous system at varying intensities. Responses were ranked from Level 1 to Level 5, corresponding to increasing degrees of observable discomfort. The reaction strength was determined by the duration of clear distress behaviors exhibited by the cricket. The images on the right illustrate representative responses: Level 1 shows minimal movement, Level 3 depicts wing movement, and Level 5 demonstrates simultaneous movement of all limbs.

Figure 2

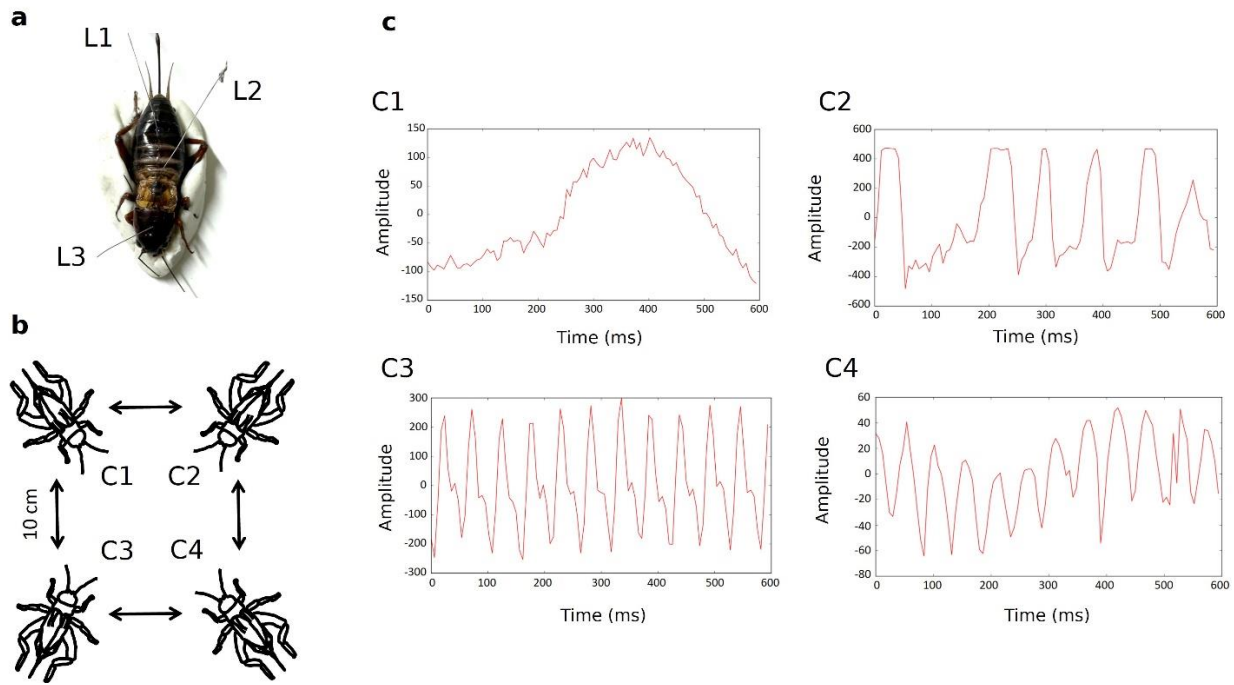

**ECG Experiment to Investigate Social Sharing of Pain in Crickets.** (a) A customized ECG probe, made from a 200-micron acupuncture needle, was inserted into the first and second-to-last exoskeletal body sections of each cricket. (b) Four crickets were arranged in a square formation, positioned 10 cm apart, facing one another. (c) Physical pain was applied to Cricket 1 (C1) by applying pressure with an object. ECG data were recorded from all crickets during the experiment. No significant heart rate changes were observed in any of the other crickets, but the muscular movements of C1 in response to pain resulted in motion artifacts in the signal.

Figure 3

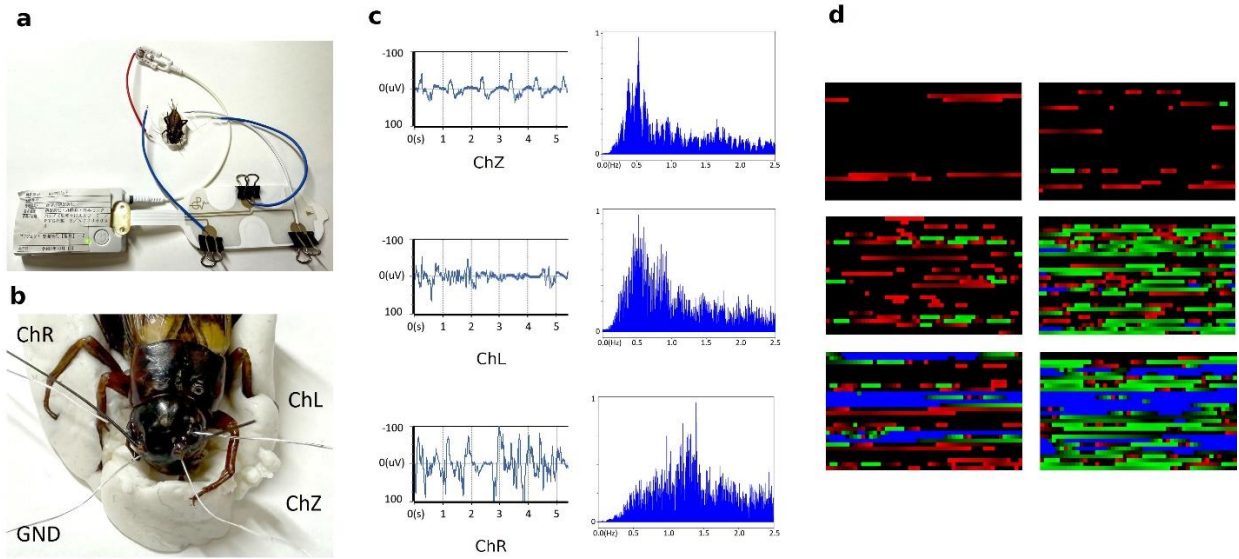

**EEG Portable Device Adapted for Cricket Nervous System Signaling.** (a) The modified EEG device is adapted for interfacing with crickets. (b) Silver wire electrodes are inserted into sensitive areas, making direct contact with the insect brain: the eyes and the base of the antennae. (c) Sample resting data from three channels are used to train the model. (d) Application of the cricket pain classifier to detect three types of stimuli: resting (red), low pain (green), and high pain (blue).

## 8. Acknowledgements

This research was supported in part by JSPS KAKENHI Grants (24K21600, 23K26069, and 23K26077). We also acknowledge the support from The Foundation for Technology Promotion of Electronic Circuit Board and Teijin Nakashima Medical Co., Ltd. Their contributions were instrumental in facilitating this study.

## 9. Data availability

Source data are provided with this paper.

## 10. Ethics declarations

### Competing interests

The authors declare no competing interests.

## 11. Contributions

M. J. M. M. and S.U. conceived of the main idea. M. J. M. M. designed the experiments. P.M. modified the readout devices and conducted the experiments. M. J. M. M analyzed the data. M. J. M. M. wrote the manuscript. S.U, KH, M.S. and R.O. revised the document. S.U. and K.H. discussed the development and supervised the project.